\documentclass[12pt,a4paper]{article}
\usepackage{a4wide}
\usepackage{amsmath}
\usepackage{amssymb}
\usepackage{epsfig}
\usepackage{subfigure}
\usepackage{exscale}
\usepackage{float}
\usepackage{bbm}
\usepackage[numbers,sort&compress]{natbib}
\usepackage{pst-plot, pstricks,pst-math}
\usepackage{fancybox,amssymb,color}
\usepackage{graphicx}
\usepackage{pstricks, color, graphicx, epsfig, psfrag}
\usepackage{amsfonts,amsmath,amssymb,slashed}
\usepackage{dsfont}
\usepackage{bbm,bm}
\usepackage{fancyhdr, a4wide}
\usepackage[english]{babel}
\usepackage{subfigure}

\makeatletter
\@addtoreset{equation}{section}
\makeatother

\title{Magnetic Catalysis in Antiferromagnetic Films}

\author{Christoph P.\ Hofmann$^a$ \\ \\
\normalsize{$^a$ Facultad de Ciencias, Universidad de Colima} \\
\vspace{0.3cm}
\normalsize{Bernal D\'iaz del Castillo 340, Colima C.P.\ 28045, Mexico} \\}

\begin{document}

\maketitle

\begin{abstract} \normalsize

We study the low-temperature behavior of antiferromagnets in two spatial dimensions that are subjected to a magnetic field oriented
perpendicular to the staggered magnetization order parameter. The evaluation of the partition function is carried to two-loop order within
the systematic effective Lagrangian technique. Low-temperature series that are valid in weak magnetic and staggered fields are derived for
the pressure, staggered magnetization, and magnetization. Remarkably, at $T$=0, the staggered magnetization is enhanced by the magnetic
field, implying that the phenomenon of magnetic catalysis also emerges in antiferromagnetic films.

\end{abstract}


\maketitle

\section{Motivation}
\label{Intro}

The thermodynamic properties of antiferromagnets in two spatial dimensions have been the topic of numerous studies. Within microscopic,
phenomenological, and numerical approaches, the free energy density, staggered magnetization, and other observables have been explored
extensively at low temperatures \citep{CHN88,AA88a,AA88b,CHN89,Tak89,GJN89,Bar91,Man91,CSY94,WY94,San97,CTVV97,BBGW98,KT98, Liu99,MS00,
SS01} -- in particular also in magnetic fields \citep{AUW77,FKLM92,Glu93,MG94,SSS94,WLW94,ZN98a,ZN98b,San99,ZC99,SM01,SVB01,Hon01,Yun02,
SS02,HSR04,VCC05,Che05,SSKPWLB05,JBMD06,HW06,HSSK07,YK07,KHK07,TZSS08,KSHK08,Syl08,CZ09,LL09,FZSR09,HRO10,MZC10,SST11,SST13,PR15}. Still,
a fully systematic analysis of how a weak magnetic field, in presence of a weak staggered field, affects the low-energy physics of
antiferromagnetic films -- both at $T$=0 and finite temperature -- appears to be lacking.

Instead of relying on phenomenological or microscopic techniques such as modified spin-wave theory, the present analysis is based on the
effective Lagrangian method that has the virtue of being fully systematic. The crucial point is that the relevant degrees of freedom in an
antiferromagnet at low temperatures -- the spin waves or magnons -- are the Goldstone bosons of a spontaneously broken internal symmetry:
$O(3) \to O(2)$.\footnote{Strictly speaking, at finite temperature and in two spatial dimensions, spontaneous symmetry breaking does not
occur because of the Mermin-Wagner theorem \citep{MW66}. However, the low-temperature physics is still dominated by the spin waves and the
staggered magnetization is different from zero at low $T$ and weak fields. In this sense the staggered magnetization is still referred to
as order parameter in the present study.} Goldstone boson effective field theory has been developed in the eighties in the context of
quantum chromodynamics \citep{GL84,GL85}, but the same universal principles can be applied to condensed matter systems
\citep{Leu94a,ABHV14}, where the phenomenon of spontaneous symmetry breaking is ubiquitous.

Within effective field theory, the thermodynamic properties of antiferromagnets in two spatial dimensions have been analyzed in
Refs.~\citep{HL90,HN93,Hof10,Hof16a}. Some of these studies -- apart from the inclusion of a staggered field -- also consider the effect
of an external magnetic field. However a systematic discussion of how the thermodynamic variables and the physics at $T$=0 depend on these
fields, has not yet been presented. In particular, the situation where the magnetic field is oriented {\it perpendicular} to the staggered
field, has not been discussed on the effective level so far. This motivates the present work where we systematically investigate the
impact of a perpendicular magnetic field onto the low-energy behavior of $d$=2+1 antiferromagnets. We evaluate the partition function up
to two-loop order, derive the low-temperature series for the free energy density, pressure, staggered magnetization, and magnetization,
and also consider the behavior of the system at zero temperature.

In the absence of a magnetic field, the spin-wave interaction does not yet manifest itself: up to two-loop order, the low-temperature
series just correspond to the free magnon gas. In nonzero magnetic fields, however, the spin-wave interaction leads to interesting
effects. In the pressure -- irrespective of the strength of the magnetic and staggered field -- the interaction among the magnons is
repulsive. Regarding the order parameter at finite temperature\footnote{See footnote 1.}, we also observe subtle effects: if the
temperature is raised from $T$=0 to finite $T$ -- while keeping the strength of the staggered and magnetic field fixed -- the order
parameter decreases as a consequence of the spin-wave interaction. Remarkably, at zero temperature, the staggered magnetization is
enhanced in weak magnetic fields. This phenomenon -- magnetic catalysis -- has been observed in quantum chromodynamics, graphene,
topological insulators, and other systems.\footnote{As we comment in subsection \ref{stagMag}, magnetic catalysis in antiferromagnetic
films is different because no charged particles or Landau levels are involved.} Finally, the perpendicular magnetic field -- both at zero
and finite temperature -- causes the magnetization to take positive values, signaling that the spins get tilted into the magnetic field
direction.

The article is organized as follows. The incorporation of the perpendicular magnetic field, along with some essential information on the
effective Lagrangian method, is discussed in Sec.~\ref{MicroEff}. The perturbative evaluation of the free energy density up to two-loop
order is provided in Sec.~\ref{Evaluation}. The low-temperature series for various thermodynamic quantities -- pressure, staggered
magnetization, and magnetization -- in presence of weak staggered and magnetic fields, are derived in Sec.~\ref{LowTSeries}. The role of
the spin-wave interaction in these observables is illustrated in various figures. In the same section we also consider the behavior at
$T$=0 and discuss the phenomenon of magnetic catalysis. Finally, in Sec.~\ref{conclusions} we present our conclusions. Technical details
on vertices with an odd number of magnon lines and the evaluation of a specific two-loop diagram can be found in two separate appendices.

\section{Microscopic and Effective Description}
\label{MicroEff}

Antiferromagnets in two spatial dimensions are described by the quantum Heisenberg model,
\begin{equation}
\label{Heisenberg}
{\cal H}_0 \ = \ - \, J \, \sum_{n.n.} {\vec S}_m \! \cdot {\vec S}_n \, ,
\qquad \qquad \qquad J = const. ,
\end{equation}
where the summation extends over all nearest neighbor spins on a bipartite lattice, and the exchange integral $J$ is negative. The
Heisenberg Hamiltonian is invariant under global internal O(3) symmetry. The antiferromagnetic ground state, however, is only invariant
under O(2). As a consequence of the spontaneously broken rotation symmetry, two spin-wave branches -- or two magnon particles -- emerge in
the low-energy spectrum. If the O(3) symmetry is exact, the two degenerate spin-wave branches follow the dispersion law
\begin{equation}
\label{disprelAF}
\omega(\vec k) \, = \, v|{\vec k}| + {\cal O}({\vec k}^3) \, , \qquad {\vec k} = (k_1,k_2) \, ,
\end{equation}
with $v$ as spin-wave velocity. According to Goldstone's theorem, both excitations obey
\begin{equation}
\lim_{{\vec k} \to 0} \omega(\vec k) = 0 \, .
\end{equation}
The symmetric model can be extended by incorporating a staggered field ${\vec H}_s$ and a magnetic field ${\vec H}$,
\begin{equation}
\label{ZeemanH}
{\cal H} = {\cal H}_0 - \sum_n {\vec S}_n \cdot {\vec H} - \sum_n (-1)^n {\vec S}_n \! \cdot {\vec H_s} \, ,
\end{equation}
that both explicitly break O(3)-invariance. Now the spontaneously broken symmetry is only approximate: the spin-wave branches exhibit an
energy gap, i.e., the magnons are no longer Goldstone bosons as they become massive. In particle physics it is common to call such
excitations {\it pseudo-Goldstone bosons}.

Let us turn to the effective description of the $d$=2+1 antiferromagnet in presence of staggered and magnetic fields. This situation has
been discussed in detail in sections IX-XI of Ref.~\citep{Hof99a} (see also Ref.~\citep{Leu94a}). Here we merely list the relevant
expressions. The basic low-energy degrees of freedom -- the two antiferromagnetic magnon fields -- we denote by $U^a = (U^1,U^2)$, and
collect them in a unit vector $U^i$,
\begin{equation}
U^i = (U^0, U^a) \, , \quad U^0 = \sqrt{1 - U^a U^a} \, ,\qquad a = 1,2 \, , \quad i = 0,1,2 \, .
\end{equation}
The ground state of the antiferromagnet is represented by ${\vec U}_0 = (1,0,0)$, and the magnons correspond to fluctuations in the
orthogonal directions.

The low-energy effective theory is based on a systematic expansion in powers of momenta, i.e., on a derivative expansion of the effective
Lagrangian. The leading term -- ${\cal L}^2_{eff}$ -- contains two space-time derivatives,
\begin{equation}
{\cal L}^2_{eff} = \mbox{$ \frac{1}{2}$} F^2 D_{\mu} U^i D^{\mu} U^i + \Sigma_s H^i_s U^i \, ,
\end{equation}
where the covariant derivative is
\begin{equation}
D_0 U^i = {\partial}_0 U^i + {\varepsilon}_{ijk} H^j U^k \, , \qquad D_r U^i = {\partial}_r U^i \qquad (r=1,2) \, .
\end{equation}
The magnetic field $H^i$ is incorporated through the time component of the covariant derivative $D_0 U^i$. On the other hand, the
staggered field $H^i_s$ couples to $\Sigma_s$ that represents the order parameter: the staggered magnetization at zero temperature, zero
external fields, and infinite volume. Apart from $\Sigma_s$, a second low-energy effective constant appears in ${\cal L}^2_{eff}$: the
quantity $F$ that is related to the spin stiffness $\rho$ (or helicity modulus) by $\rho = F^2$. Note that the magnetic field counts as
order $p$ like the time derivative, whereas the staggered field is of order $p^2$.

The subleading piece in the effective Lagrangian is of order $p^4$, 
\begin{eqnarray}
\label{Leff4}
{\cal L}^4_{eff} & = & e_1 (D_{\mu} U^i D^{\mu} U^i)^2 + e_2 (D_{\mu} U^i D^{\nu} U^i)^2
+ k_1 \frac{\Sigma_s}{F^2} (H_s^i U^i) (D_{\mu} U^k D^{\mu} U^k) \nonumber \\
& & + k_2 \frac{{\Sigma}_s^2}{F^4} (H_s^i U^i)^2 + k_3 \frac{{\Sigma}_s^2}{F^4} H_s^i H_s^i \, ,
\end{eqnarray}
and contains five next-to-leading order (NLO) effective constants whose numerical values have to be determined or estimated to make the
effective field theory predictive (see below).

We now comment on an important issue related to Lorentz-invariance. The leading and next-to-leading effective Lagrangians are
Lorentz-invariant. In view of the fact that the underlying bipartite lattices are not even space-rotation invariant, why is our approach
legitimate? The first observation is that the leading piece ${\cal L}^2_{eff}$ is strictly (pseudo-)Lorentz-invariant, the spin-wave
velocity $v$ taking the role of the speed of light. This accidental symmetry emerges because lattice anisotropies only show up at order
$p^4$ (and beyond) in the effective Lagrangian \citep{HN93}. On the other hand, in ${\cal L}^4_{eff}$ one should include all additional
terms that are permitted by the lattice geometry. However, as we explain below, these effects only start manifesting themselves at
next-to-next-to leading order in the low-temperature expansion which is beyond two-loop accuracy we pursue in the present evaluation. This
perfectly justifies maintaining a (pseudo-)Lorentz-invariant structure also in ${\cal L}^4_{eff}$.

In the following we consider the scenario where the magnetic field points into a direction perpendicular to the staggered field,
\begin{equation}
\label{externalFields}
{\vec H}_{\perp} = (0,H,0) \, , \qquad {\vec H}_s = (H_s,0,0) \, .
\end{equation}
While we have chosen ${\vec H}_{\perp}$ to point into the 1-direction, the physics would be the same had we chosen the 2-direction. Note
that the staggered field points into the 0-direction, aligned with the staggered magnetization order parameter or ground state
${\vec U}_0 = (1,0,0)$.

The leading-order effective Lagrangian ${\cal L}^2_{eff}$ gives rise to the following magnon dispersion relations,
\begin{eqnarray}
\label{disprelAFH}
\omega_{I} & = & \sqrt{{\vec k}^2 + \frac{\Sigma_s H_s}{F^2} + H^2} \, , \nonumber \\
\omega_{I\!I} & = & \sqrt{{\vec k}^2 + \frac{\Sigma_s H_s}{F^2}} \, .
\end{eqnarray}
These results coincide with the expressions derived within microscopic or phenomenological descriptions -- see, e.g.,
Refs.~\citep{KT58,ABK61a}. Remarkably, one of the magnons is not affected by the magnetic field. The structure of the dispersion relation
is relativistic in both cases, and the corresponding "magnon mass terms" are identified as
\begin{equation}
\label{masses}
M^2_{I} = \frac{\Sigma_s H_s}{F^2} + H^2 \, , \qquad M^2_{I\!I} = \frac{\Sigma_s H_s}{F^2} \, .
\end{equation}
Note that the staggered field emerges linearly while the dependence on the magnetic field is quadratic. If the fields are switched off, we
reproduce the linear and ungapped dispersion relation Eq.~(\ref{disprelAF}).\footnote{We have set the spin-wave velocity $v$ to one.}

It is convenient to utilize dimensional regularization in the perturbative evaluation of the partition function. The zero-temperature
propagators for antiferromagnetic magnons, in presence of ${\vec H}_s=(H_s,0,0)$ and ${\vec H}_{\perp}=(0,H,0)$, amount to
\begin{equation}
\label{regprop}
\Delta^{I,I\!I} (x) = (2 \pi)^{-d} \int {\mbox{d}}^d p \, e^{ipx} (M_{I,I\!I}^2 + p^2)^{-1}
= {\int}_{\!\!\!0}^{\infty} \mbox{d} \rho \, (4 \pi \rho)^{-d/2} e^{-\rho M_{I,I\!I}^2 - x^2/{4 \rho}} \, ,
\end{equation}
where $M_{I}$ and $M_{I\!I}$ are defined in Eq.~(\ref{masses}). The corresponding thermal propagators in Euclidean space are given by
\begin{equation}
\label{ThermalPropagator}
G^{I,I\!I}(x) = \sum_{n = - \infty}^{\infty} \Delta^{I,I\!I}({\vec x}, x_0 + n \beta) \, , \qquad \beta = \frac{1}{T} \, .
\end{equation}

We emphasize that the magnetic and staggered field are treated as perturbations that explicitly break O(3) invariance of the
Heisenberg Hamiltonian. As long as these fields are weak, the O(3) symmetry still is approximate, and our basic setting is valid: it is
conceptually consistent to start from the collinear antiferromagnetic ground state and interpret the two magnons as oscillations of the
staggered magnetization order parameter. It is well-known, however, that in presence of magnetic fields perpendicular to the
staggered magnetization, the spins get tilted, creating a canted (non-collinear) phase (see, e.g., Refs.~\citep{WLW94,KSHK08,MZC10}). If
the canting angle is large, the magnetic field can no longer be considered as a small perturbation. Rather the canted phase
should be chosen as the starting configuration underlying the perturbative expansion. Most importantly, since the spontaneous symmetry
breaking pattern then is O(3) $\to$ 1, not two, but three Goldstone fields emerge in the low-energy spectrum. This scenario, i.e., the
low-temperature physics of canted phases, we postpone for future studies -- in the present investigation we consider weak magnetic fields
where only two spin-wave branches are relevant.

\section{Evaluation of the Partition Function}
\label{Evaluation}

\begin{figure}
\begin{center}
\includegraphics[width=15cm]{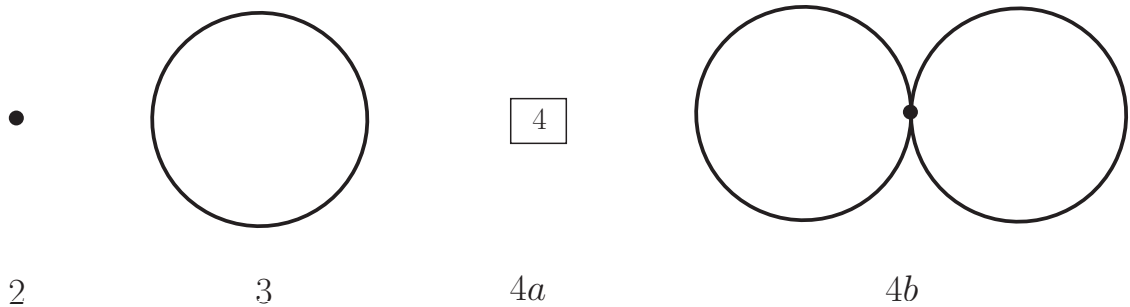}
\end{center}
\caption{Low-temperature expansion of the partition function for the $d$=2+1 antiferromagnet: Feynman diagrams up to two-loop order $T^4$.
Filled circles refer to ${\cal L}^2_{eff}$, while the vertex associated with the subleading piece ${\cal L}^4_{eff}$ is represented by the
number 4. Each loop is suppressed by one power of $T$.}
\label{figure1}
\end{figure}

We now evaluate the partition function for the $d$=2+1 antiferromagnet subjected to the magnetic and staggered fields defined in
Eq.~(\ref{externalFields}). The relevant Feynman diagrams up to two-loop order are shown in Fig.~\ref{figure1}.\footnote{The perturbative
evaluation of the partition function is described in more detail in section 2 of Ref.~\citep{Hof10} and in appendix A of
Ref.~\citep{Hof11}. Regarding the effective Lagrangian technique in general, the interested reader may consult
Refs.~\citep{Leu95,Sch03,Bra10}.} The crucial point is that we are dealing with a systematic low-temperature expansion of the partition
function where each magnon loop is suppressed by one power of temperature. The free Bose gas contribution is given by the one-loop graph
$3$ (order $T^3$), while the two-loop graph $4b$ is of order $T^4$.

\begin{figure}
\begin{center}
\includegraphics[width=15cm]{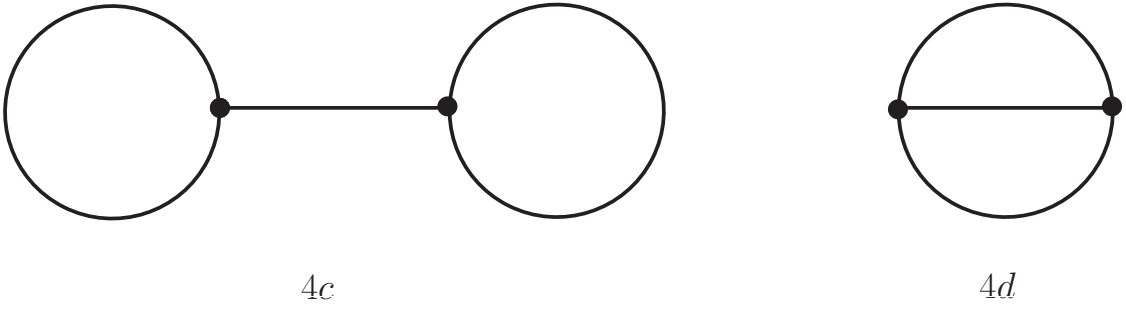}
\end{center}
\caption{Low-temperature expansion of the partition function for the $d$=2+1 antiferromagnet: additional Feynman diagrams up to two-loop
order $T^4$ emerging in presence of a perpendicular magnetic field. Filled circles refer to ${\cal L}^2_{eff}$. Each loop is suppressed by
one power of $T$.}
\label{figure2}
\end{figure}

The incorporation of a perpendicular magnetic field generates extra vertices that involve an {\it odd} number of magnon lines. With
respect to ${\cal L}^2_{eff}$, the explicit terms are proportional to one time derivative and read
\begin{equation}
i F^2 H \, \Big( U^0 \partial_0 U^2 - U^2 \partial_0 U^0 \Big) \, .
\end{equation}
These contributions, along with those originating from ${\cal L}^4_{eff}$, create vertices with $1,3,5, \dots$ magnon lines: in presence of
a perpendicular magnetic field, the set of Feynman diagrams has to be extended by the graphs depicted in Fig.~\ref{figure2}. Note that in
the diagrams depicted in Fig.~\ref{figure1}, the magnetic field manifests itself implicitly in the thermal propagator $G^I(x)$ through
$M_{I}$.

The tree graphs $2$ and $4a$ merely contribute to the energy density at zero temperature,
\begin{eqnarray}
\label{z2z4a}
z_2 & = & - \Sigma_s H_s - \mbox{$ \frac{1}{2}$} F^2 H^2 \, , \nonumber \\
z_{4a} & = & -(k_2 + k_3) \frac{\Sigma^2_s H^2_s}{F^4} - k_1 \frac{\Sigma_s H_s}{F^2} H^2 -(e_1 + e_2) H^4 \, .
\end{eqnarray}
The dominant temperature-dependent contribution comes from one-loop graph $3$,
\begin{eqnarray}
z_3 & = & - \mbox{$ \frac{1}{2}$} {(4 \pi)}^{-d/2} \Gamma(-\mbox{$ \frac{d}{2}$}) \Big\{ M_{I}^d + M_{I\!I}^d \Big\}
- \mbox{$ \frac{1}{2}$} \Big\{ g^{I}_0 + g^{I\!I}_0 \Big\} \nonumber \\
& = & - \mbox{$ \frac{1}{2}$} \Big\{ g^{I}_0 + g^{I\!I}_0 \Big\} - \frac{1}{12 \pi}
\Bigg\{ {\Big( \frac{\Sigma_s H_s}{F^2} + H^2 \Big) }^{3/2} + {\Big( \frac{\Sigma_s H_s}{F^2} \Big)}^{3/2} \Bigg\} \, .
\end{eqnarray}
Note that the Gamma function is finite in two spatial dimensions,
\begin{equation}
\lim_{d \to 3} \, \mbox{$ \frac{1}{2}$} {(4 \pi)}^{-d/2} \Gamma(-\mbox{$ \frac{d}{2}$}) = \frac{1}{12 \pi} \, .
\end{equation}
The quantities $g^{I}_0$ and $g^{I\!I}_0$ are the kinematical functions related to the free magnon gas,
\begin{equation}
\label{BoseFunctions}
g^{I,{I\!I}}_r(H_s, H, T) = 2 {\int}_{\!\!\! 0}^{\infty} \frac{\mbox{d} \rho}{(4 \pi \rho)^{d/2}} \, {\rho}^{r-1} \, 
\exp\Big(- \rho M_{I,I\!I}^2 \Big) \, \sum_{n=1}^{\infty} \exp(- n^2/{4 \rho T^2}) \, .
\end{equation}
Next, the two-loop graph $4b$ contributes with
\begin{eqnarray}
z_{4b} & = & - \frac{\Sigma_s H_s}{8 F^4} \, {\Big( G^{I}_1 - G^{I\!I}_1 \Big)}^2 - \frac{H^2}{2 F^2} \, {\Big( G^{I}_1 \Big)}^2 \nonumber \\
& = & -\frac{\Sigma_s H_s}{8 F^4} \, \Big\{ {(g^{I}_1)}^2 - 2 g^{I}_1 g^{I\!I}_1 + {(g^{I\!I}_1)}^2 \Big\} - \frac{H^2}{2 F^2}{(g^{I}_1)}^2
\nonumber \\
& & + \frac{\Sigma_s H_s}{16 \pi F^4} \, \Bigg\{ \sqrt{\frac{\Sigma_s H_s}{F^2} + H^2} - \frac{\sqrt{\Sigma_s H_s}}{F} \Bigg\} g^{I}_1
+ \frac{H^2}{4 \pi F^2} \, \sqrt{\frac{\Sigma_s H_s}{F^2} + H^2} \, g^{I}_1 \nonumber \\
& & - \frac{\Sigma_s H_s}{16 \pi F^4} \, \Bigg\{ \sqrt{\frac{\Sigma_s H_s}{F^2} + H^2} - \frac{\sqrt{\Sigma_s H_s}}{F} \Bigg\} g^{I\!I}_1
-\frac{\Sigma_s^2 H_s^2}{64 \pi^2 F^6} - \frac{5 \Sigma_s H_s H^2}{128 \pi^2 F^4} \nonumber \\
& & - \frac{H^4}{32 \pi^2 F^2} + \frac{\Sigma_s^{3/2} H_s^{3/2}}{64 \pi^2 F^5} \, \sqrt{\frac{\Sigma_s H_s}{F^2} + H^2} \, ,
\end{eqnarray}
where $G^{I,I\!I}_1$ are the thermal propagators evaluated at the origin,
\begin{equation}
\label{defG1}
G^{I,I\!I}_1 \, = \, {G^{I,I\!I}(x)}_{|x=0} \, = \, g^{I,I\!I}_1 - \frac{M_{I,I\!I}}{4 \pi} \, .
\end{equation}
In the absence of the magnetic field, we have $G^{I}_1 = G^{I\!I}_1$, such that the entire two-loop contribution vanishes.

\begin{figure}
\begin{center}
\includegraphics[width=10.5cm]{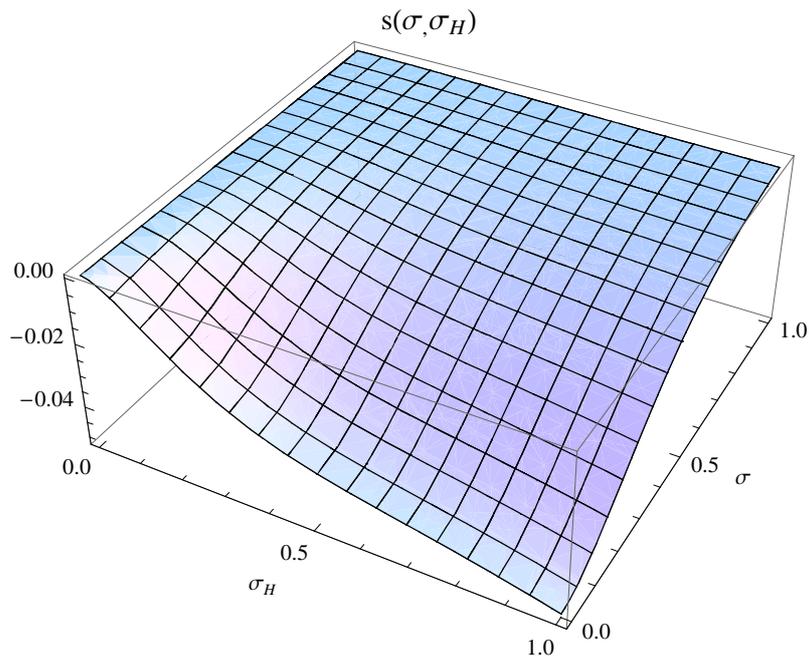}
\end{center}
\caption{[Color online] The function $s(\sigma,\sigma_H)$, where $\sigma$ and $\sigma_H$ are the dimensionless parameters
$\sigma=\sqrt{\Sigma_s H_s}/(2 \pi F T)$ and $\sigma_H=H/(2 \pi T)$.}
\label{figureSunny}
\end{figure}

Finally, the explicit evaluation of diagram $4c$ yields zero,
\begin{equation}
z_{4c} =0 \, ,
\end{equation}
while the sunset diagram amounts to 
\begin{equation}
z_{4d} = \frac{2 H^2}{F^2} \, \int_{\cal T} {\mbox{d}}^d x \, G^{I}(x) \, \partial_0 G^{I}(x) \, \partial_0 G^{I\!I}(x) \, .
\end{equation}
This integral over the torus ${\cal T} = {\cal R}^{d_s} \times S^1$, with circle $S^1$ defined as $- \beta / 2 \leq x_0 \leq \beta / 2$, is
divergent in the ultraviolet. The renormalization of this expression and the evaluation of the thermal sums is described in appendix
\ref{appendixB}. The finite contribution to the free energy density is given by
\begin{equation}
{\overline z_{4d}} = \frac{2}{F^2} \, s(\sigma,\sigma_H) \, T^4 \, .
\end{equation}
The dimensionless function $s(\sigma,\sigma_H)$ is defined in Eq.~(\ref{sunny}), and the dimensionless parameters $\sigma$ and $\sigma_H$
are
\begin{equation}
\sigma = \frac{\sqrt{\Sigma_s H_s}}{2 \pi F T} = m \frac{F^2}{T} \, , \qquad \sigma_H = \frac{H}{2 \pi T} = m_H \frac{F^2}{T} \, .
\end{equation}
A plot of $s(\sigma,\sigma_H)$ is provided in Fig.~\ref{figureSunny}.

Remarkably, up to two-loop order, the NLO effective constants $e_1, e_2, k_1, k_2, k_3$ only show up in the tree graph $4a$. These
constants -- that are a priori unknown -- hence only matter in temperature-independent contributions. The low-temperature expansion, in
particular the impact of the spin-wave interaction, is governed by the leading effective Lagrangian ${\cal L}^2_{eff}$. Up to two-loop
order, the thermal properties of the $d$=2+1 antiferromagnet are thus rigorously captured by our effective field theory approach that is
based on (pseudo-)Lorentz invariance. The specific geometry of the underlying bipartite lattice is irrelevant as far as the structure of
the low-temperature expansion is concerned. Alternatively, this can be seen as follows. Lattice anisotropies modify the dispersion
relation
\begin{equation}
\omega(\vec k) \, = \, v|{\vec k}| + {\cal O}({\vec k}^3)
\end{equation}
at order ${\vec k}^3$ -- the specific terms and coefficients indeed depend on the lattice geometry. While the linear term in the
dispersion relation yields the dominant contribution of order $T^3$ in the free energy density, the corrections $\propto {\vec k}^3$
contribute at order $T^5$ which is beyond our scope. Therefore our (pseudo-)Lorentz-invariant framework is perfectly legitimate: we make
no mistake by merely considering the leading term in the dispersion relation.

The lattice structure only reflects itself in the numerical values of the leading-order effective constants $F$ and $\Sigma_s$ that have
been determined with high-precision loop-cluster algorithms. For the square lattice \citep{GHJNW09} they read
\begin{equation}
\rho = 0.1808(4) J \, , \quad \Sigma_s = 0.30743(1) / a^2 \, , \quad v = 1.6585(10) J a \qquad (S=\mbox{$ \frac{1}{2}$}) \, ,
\end{equation}
for the honeycomb lattice \citep{JKNW08} they are
\begin{equation}
\rho = 0.102(2) J \, , \quad {\tilde \Sigma_s} = 0.2688(3) \, , \quad v = 1.297(16) J a \qquad (S=\mbox{$ \frac{1}{2}$}) \, ,
\end{equation}
with
\begin{equation}
{\tilde \Sigma_s} = \frac{ 3 \sqrt{3}}{4} \, \Sigma_s \, a^2.
\end{equation}
Note that the spin stiffness $\rho$ as well as $\Sigma_s$, much like the spin-wave velocity $v$, are given in units of $J$ (exchange
integral) and $a$ (lattice size).

\section{Low-Temperature Series}
\label{LowTSeries}

The low-temperature physics of the system can be captured by various dimensionless ratios. As independent quantities we define the
parameters $m, m_H$ and $t$ as
\begin{equation}
\label{definitionRatios}
m \equiv \frac{\sqrt{\Sigma_s H_s}}{2 \pi F^3} \, , \qquad
m_H \equiv \frac{H}{2 \pi F^2} \, \, , \qquad
t \equiv \frac{T}{2 \pi F^2} \, .
\end{equation}
For the effective low-energy expansion to be consistent, the temperature as well as the staggered and magnetic field must be small
compared to the scale $\Lambda$ that characterizes the microscopic system. The natural scale in the Heisenberg antiferromagnet is the
exchange integral $J$. In the present study, we define {\it low} temperatures and {\it weak} fields by 
\begin{equation}
\label{domain}
T, \, H, \, M_{I\!I} (\propto \sqrt{H_s}) \ \lesssim 0.3 \ J \, .
\end{equation}
The factors $2 \pi$ in Eq.~(\ref{definitionRatios}) were introduced in analogy to the relevant scale in quantum chromodynamics (see
Ref.~\cite{Hof10}). The point is that for the antiferromagnet -- both on the square and honeycomb lattice -- the denominator $2 \pi F^2$
is of the order of $J$. The parameters $m, m_H, t$ hence measure temperature and field strength relative to the underlying microscopic
scale.

Whereas temperature and magnetic field can be arbitrarily small, it should be noted that the staggered field can not be switched off. This
is a consequence of the Mermin-Wagner theorem \citep{MW66} and the fact that the staggered magnetization -- unlike the magnetization --
represents the order parameter. As we have discussed on previous occasions, the domain where the effective expansion fails due to the
smallness of the staggered field, is tiny. The interested reader is referred to Figs.~2 and 3 of Ref.~\citep{Hof16a}.

\subsection{Pressure}

We first discuss the pressure, defined by
\begin{equation}
P = z_0 - z \, .
\end{equation}
The quantity $z_0$ includes all terms in the energy density that do not depend on temperature. Introducing dimensionless functions
$h_i(m,m_H,t)$ as
\begin{eqnarray}
\label{ThermalDimensionlessD2}
& & g_0(m,m_H,t) = T^3 \, h_0(m,m_H,t) , \quad g_1(m,m_H,t) = T \, h_1(m,m_H,t) \, , \nonumber \\
& & g_2(m,m_H,t) = \frac{h_2(m,m_H,t)}{T} , \quad g_3(m,m_H,t) = \frac{h_3(m,m_H,t)}{T^3} \, ,
\end{eqnarray}
the structure of the low-temperature series becomes more transparent because powers of temperature are explicit. For the pressure we get
\begin{eqnarray}
\label{pressureAF}
& & P(T,H_s,H) = {\tilde p}_1 \, T^3 + {\tilde p}_2 \, T^4 + {\cal O}(T^5) \, , \nonumber \\
& & \quad {\tilde p}_1(T,H_s,H) = \mbox{$ \frac{1}{2}$} \Big\{ h^{I}_0 + h^{I\!I}_0 \Big\} \, , \nonumber \\
& & \quad {\tilde p}_2(T,H_s,H) = \frac{m^2}{8 F^2 t^2} \, {(h^{I}_1 - h^{I\!I}_1)}^2
+ \frac{m^2_H}{2 F^2 t^2}{(h^{I}_1)}^2 \nonumber \\
& & \hspace{2.6cm} - \frac{m^2}{16 \pi F^2 t^3} \, \Bigg\{ \sqrt{m^2 + m^2_H} - m \Bigg \} \Big( h^{I}_1 - h^{I\!I}_1 \Big) \nonumber \\
& & \hspace{2.6cm} - \frac{m_H^2}{4 \pi F^2 t^3} \, \sqrt{m^2 + m^2_H} \, h^{I}_1 - \frac{2}{F^2} \, s(\sigma,\sigma_H) \, .
\end{eqnarray}
\begin{figure}
\begin{center}
~~~~\hbox{
\includegraphics[width=7.5cm]{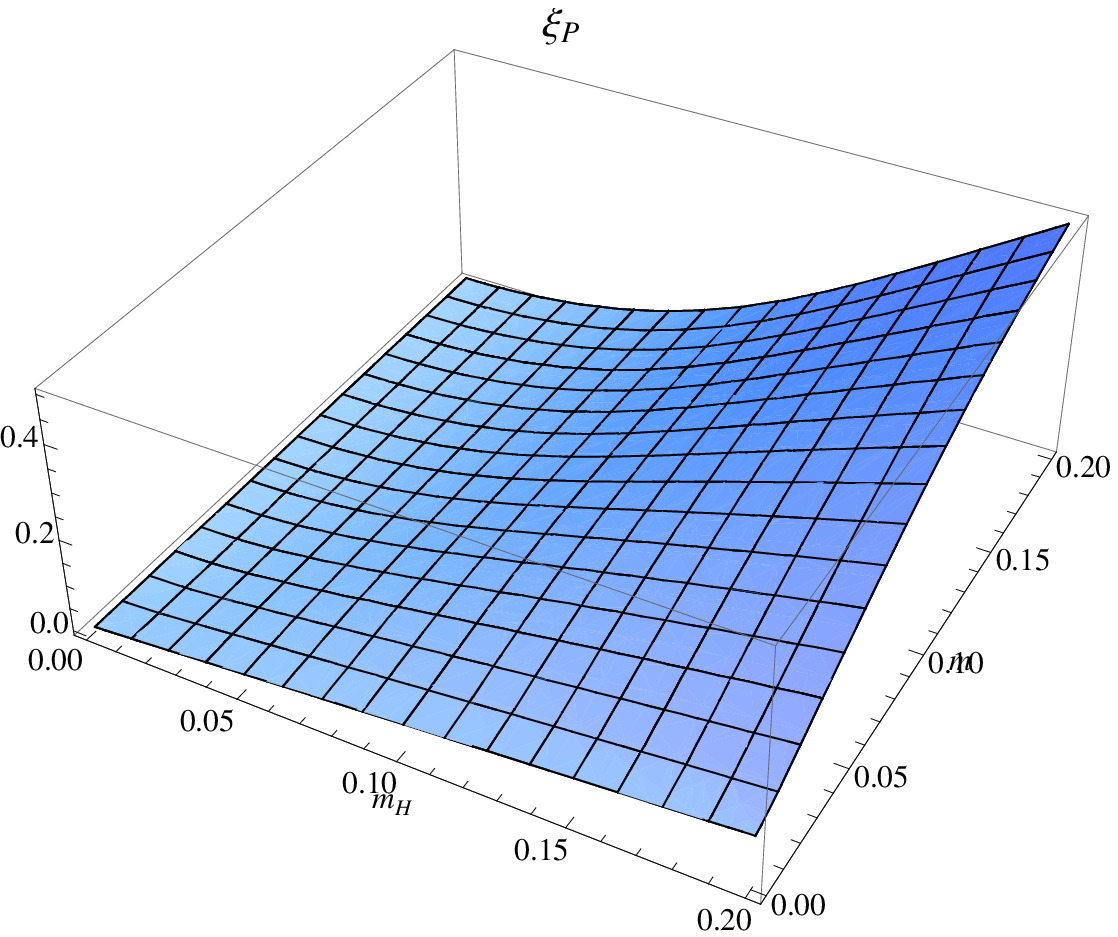}
\includegraphics[width=7.5cm]{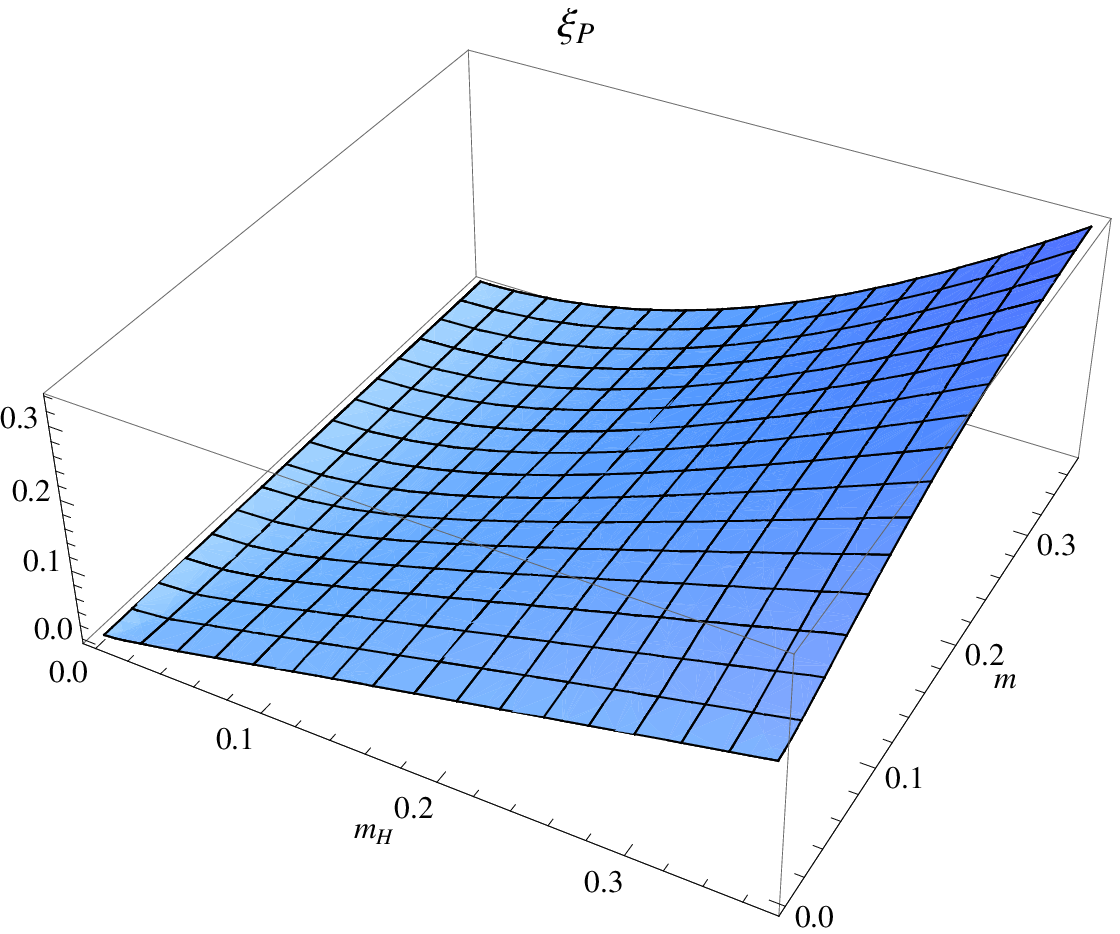}}
\end{center}
\caption{[Color online] Spin-wave interaction manifesting itself in the pressure -- measured by $\xi_P(T,H_s,H)$ -- of $d$=2+1
antiferromagnets as a function of magnetic and staggered field at the temperatures $T/2 \pi F^2 = 0.02$ (left) and $T/2 \pi F^2 = 0.1$
(right).}
\label{figure3}
\end{figure}
The dominant term of order $T^3$ (graph $3$) corresponds to the free Bose gas. The term of order $T^4$ (graphs $4b$ and $4d$) represents
the leading interaction contribution. In the absence of a perpendicular magnetic field, the spin-wave interaction does not manifest itself
at two-loop order: the corresponding coefficient ${\tilde p}_2$ is zero. On the other hand, if a perpendicular magnetic field is present,
the behavior of the system is quite interesting: in Fig.~\ref{figure3} we depict the ratio
\begin{equation}
\label{intRatioP}
\xi_P(T,H_s,H) = \frac{P_{int}(T,H_s,H)}{P_{Bose}(T,H_s,H)} = \frac{{\tilde p}_2 T^4}{{\tilde p}_1 T^3}
\end{equation}
that measures strength and sign of the spin-wave interaction in the pressure relative to the free Bose gas contribution. The plots refer
to the temperatures $T/2 \pi F^2 = 0.02$ (left) and $T/2 \pi F^2 = 0.1$ (right). As the figure suggests, irrespective of the strength of
the magnetic and staggered field, the interaction among antiferromagnetic magnons is repulsive.\footnote{It has been argued previously
that the limit $H_s \to 0$ becomes problematic in any thermodynamical observable. However, according to Ref.~\citep{Hof16a}, the error
introduced in the pressure is tiny and not visible in Fig.~\ref{figure3} of the present work.}

\subsection{Staggered Magnetization}
\label{stagMag}

The staggered magnetization order parameter can be extracted from the free energy density by
\begin{equation}
\Sigma_s(T,H_s,H) = - \frac{\partial z(T,H_s,H)}{\partial H_s} \, .
\end{equation}
The low-temperature series takes the structure\footnote{We do not display the coefficient ${\tilde \sigma}_2$ since the expression is
rather lengthy -- it can trivially be obtained from $z_{4b}$ given in Sec.~\ref{Evaluation}.}
\begin{eqnarray}
\label{OPAF}
& & \Sigma_s(T,H_s,H) = \Sigma_s(0,H_s,H) + {\tilde \sigma}_1 T + {\tilde \sigma}_2 T^2
+ {\cal O}(T^3) \, , \nonumber \\
& & \ {\tilde \sigma}_1(T,H_s,H) = -\frac{\Sigma_s}{2 F^2} \, \Big( h^{I}_1 + h^{I\!I}_1 \Big) \, .
\end{eqnarray}
The spin-wave interaction comes into play at order $T^2$. Again, in zero magnetic field, there is no interaction term at two-loop order:
${\tilde \sigma}_2(T,H_s,0) = 0$.

\begin{figure}
\begin{center}
~~~~\hbox{
\includegraphics[width=7.5cm]{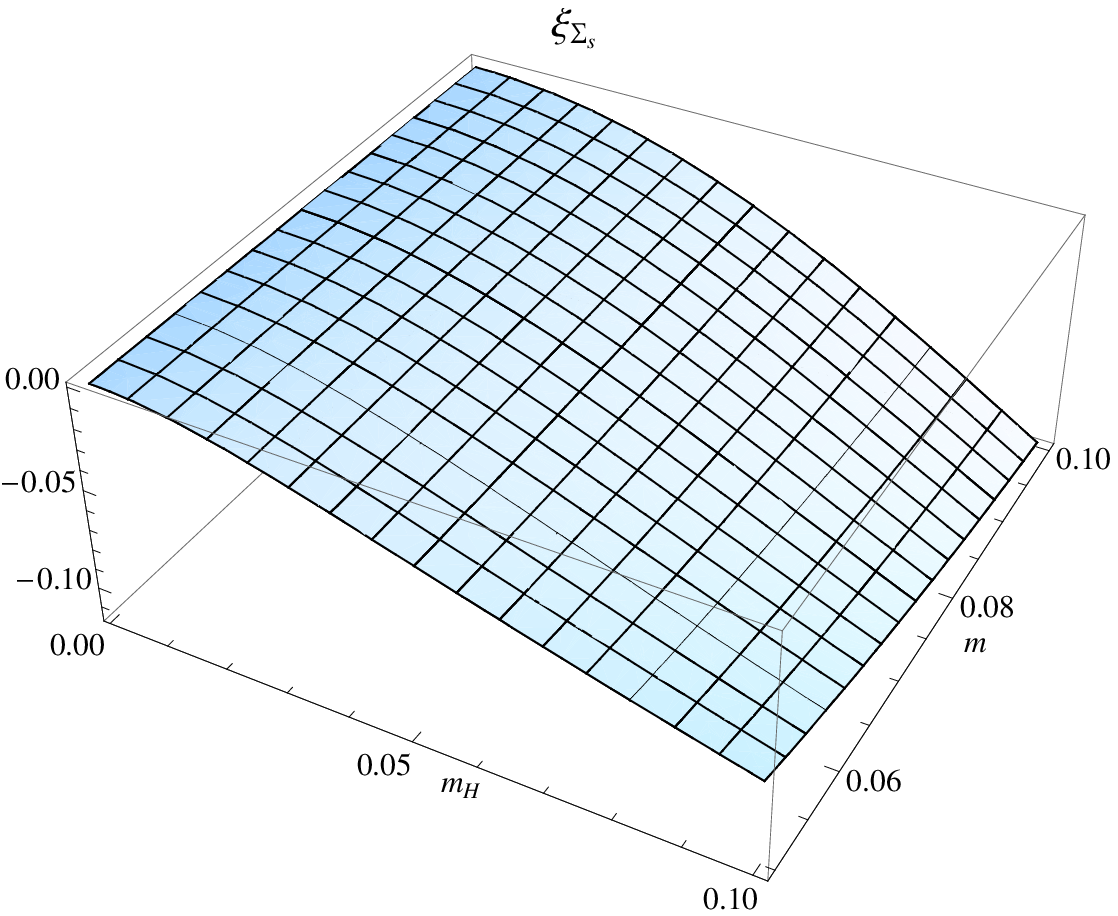}
\includegraphics[width=7.5cm]{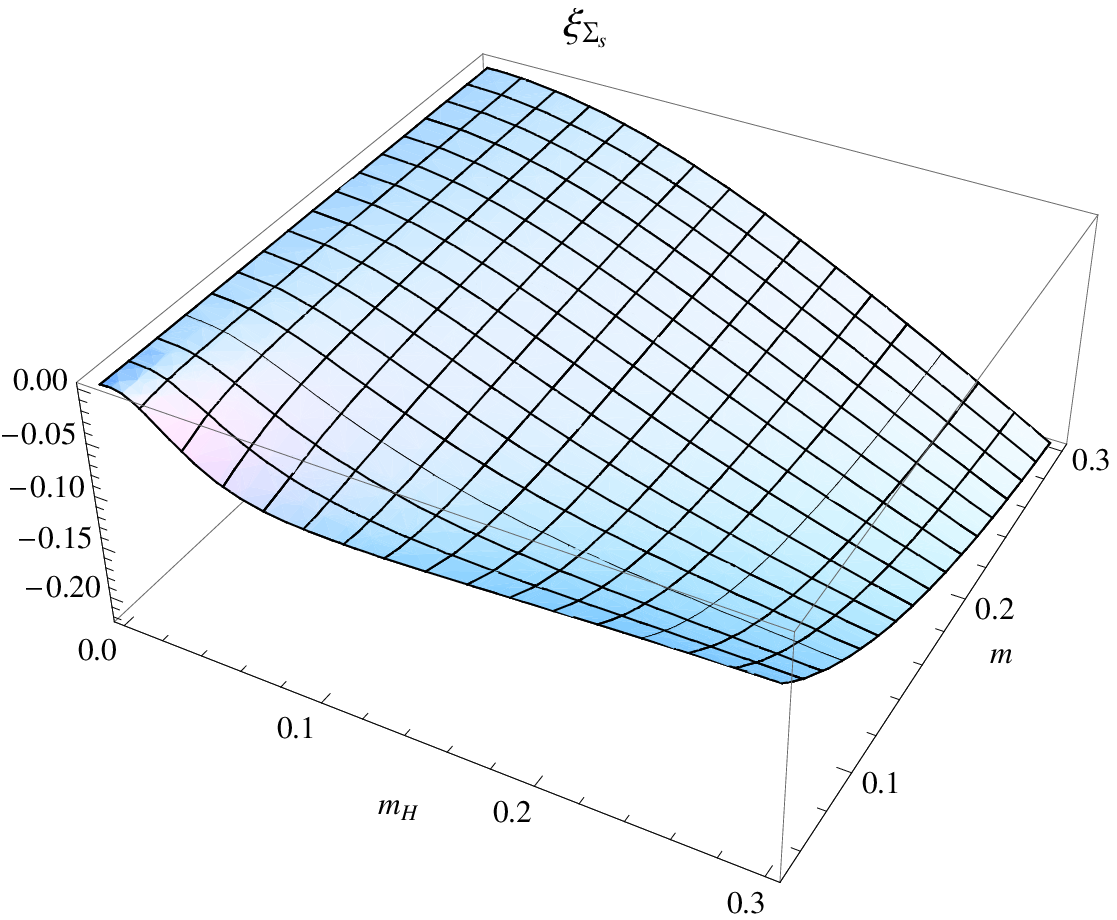}}
\end{center}
\caption{[Color online] Spin-wave interaction manifesting itself in the staggered magnetization -- measured by $\xi_{\Sigma_s}(T,H_s,H)$
-- of $d$=2+1 antiferromagnets as a function of magnetic and staggered field at the temperatures $T/2 \pi F^2 = 0.02$ (left) and
$T/2 \pi F^2 = 0.1$ (right).}
\label{figure5}
\end{figure}

To explore the impact of the spin-wave interaction in the order parameter, we consider the ratio 
\begin{equation}
\xi_{\Sigma_s}(T,H_s,H) = \frac{{\Sigma}_{s,int}(T,H_s,H)}{|{\Sigma}_{s,Bose}(T,H_s,H)|} = \frac{{\tilde \sigma}_2 T^2}{|{\tilde \sigma}_1| T}
\, ,
\end{equation}
that we depict in Fig.~\ref{figure5} for the temperatures $T/2 \pi F^2 = \{ 0.02, 0.1 \}$. The quantity $\xi_{\Sigma_s}(T,H_s,H)$ is
negative in the parameter region we consider. Negative $\xi_{\Sigma_s}$ means that if the temperature is raised from $T$=0 to finite $T$ --
while keeping $H_s$ and $H$ fixed -- the order parameter decreases due to the spin-wave interaction.

Recall that it makes no sense to address the two-dimensional system in very weak staggered fields within our framework, because the
effective expansion breaks down when one approaches the limit $H_s \to 0$.\footnote{A detailed discussion of how this relates to the
Mermin-Wagner theorem, can be found at the end of section 4 in Ref.~\citep{Hof14}.} In our plots we have chosen the staggered field
strength as
\begin{equation}
0.05 \le m \lesssim 0.3 \, , \qquad m =\frac{\sqrt{\Sigma_s H_s}}{2 \pi F^3} \, .
\end{equation}
This guarantees that the effects we observe are indeed physical and not just artifacts of our effective calculation extrapolated to a
forbidden parameter region. 

At zero temperature, the order parameter is given by
\begin{eqnarray}
\label{OPT0}
\frac{\Sigma_s(0,H_s,H)}{\Sigma_s} & = & 1 + \frac{m}{4} + \frac{\sqrt{m^2+m_H^2}}{4} + \frac{m^2}{8} + \frac{5 m_H^2}{32}
- \frac{m^3}{8 \sqrt{m^2+m_H^2}} \nonumber \\
& & - \frac{3 m \, m_H^2}{32 \sqrt{m^2+m_H^2}} + 8 \pi^2 F^2 (k_2 + k_3) \, m^2 + 4 \pi^2 F^2 k_1 \, m_H^2 \, , \nonumber \\
& & \hspace{-2.4cm} m = \frac{\sqrt{\Sigma_s H_s}}{2 \pi F^3} \, , \qquad m_H = \frac{H}{2 \pi F^2} \, , \qquad 
\Sigma_s = \Sigma_s(0,0,0) \, .
\end{eqnarray}
In contrast to finite temperature, at $T$=0, next-to-leading order effective constants arise in the low-energy expansion of the staggered
magnetization. The actual values of these constants depend on the underlying system and are not fixed by the symmetries. They should be
determined by numerical simulations, comparison with microscopic calculations, or through experiments. Unfortunately, in the case of the
$d$=2+1 antiferromagnet, none of these options seems to be available.\footnote{The exception is Ref.~\citep{GHJNW09} where the combination
$k_2+k_3$ of NLO effective constants was determined using a loop-cluster algorithm.} Still, their magnitude can be estimated. According to
Ref.~\citep{GJMM16} they are very small, of order
\begin{equation}
|k_1| \approx |k_2| \approx |k_3| \approx \frac{1}{64 \pi^3 F^2} \approx \frac{0.0005}{F^2} \, ,
\end{equation}
much like the other NLO effective constants $e_1$ and $e_2$. It should be noted that the above estimate concerns their magnitude, but
leaves open their signs. However these corrections are small -- moreover, the dominant contributions in the series (\ref{OPT0}) do not
involve NLO effective constants.

At $T$=0 and in zero magnetic field, the series is characterized by powers of $\sqrt{H_s}$, 
\begin{equation}
\label{OPinHszeroH}
\Sigma_s(0,H_s,0) = \Sigma_s + \frac{{\Sigma_s}^{3/2}}{4 \pi F^3} \, \sqrt{H_s } + \frac{2 {\Sigma_s}^2}{F^4} \, (k_2 + k_3) \, H_s +
{\cal O}(H_s^{3/2}) \, ,
\end{equation}
and in zero staggered field\footnote{It is perfectly legitimate at $T$=0 to consider the limit $H_s \to 0$. Only at finite $T$ it is
inconsistent to switch off the staggered field in our effective field theory approach.} by powers of $H$,
\begin{equation}
\label{OPinHzeroHs}
\Sigma_s(0,0,H) = \Sigma_s + \frac{\Sigma_s}{8 \pi F^2} \, H + \frac{\Sigma_s}{F^2} \,
\Big\{ k_1 + \frac{5}{128 \pi^2 F^2} \Big\} \, H^2 + {\cal O}(H^3) \, .
\end{equation}
While the order parameter is indeed expected to increase when the staggered field becomes stronger, the behavior with respect to the
magnetic field comes rather unexpectedly: in the series (\ref{OPinHzeroHs}), the term linear in $H$ is small, but positive. The order
parameter thus increases when a weak perpendicular magnetic field is applied. Notice that the subleading correction (order $H^2$) involves
the NLO effective constant $k_1$ whose sign remains open. Still, the behavior of the order parameter in weak magnetic fields is dominated
by the leading term that is strictly positive. We emphasize that this result is universal in the sense that the term of order $H$ is the
same for any bipartite lattice: the only difference between, e.g., the square and honeycomb lattice, concerns the actual values of the
effective constants $\Sigma_s$ and $F$.

The phenomenon that the order parameter is enhanced by an external magnetic field when the order parameter is already present in zero
magnetic field, is called {\bf magnetic catalysis} according to Ref.~\citep{ANT16}. It has been observed in quantum chromodynamics, where
the quark condensate -- the order parameter of the spontaneously broken chiral symmetry -- increases at $T$=0 in presence of a magnetic
field \citep{SS97,MS02,Oza14,ANT16}. Magnetic catalysis has also been reported in condensed matter systems like graphene \citep{Sho13} and
three-dimensional topological insulators \citep{GMSS16}. The fact that the staggered magnetization is enhanced at $T$=0 in square lattice
antiferromagnets subjected to a magnetic field perpendicular to the order parameter, has been reported in Ref.~\citep{LL09}.

It should be pointed out that the phenomenon of magnetic catalysis -- as it is discussed in the context of QCD or, e.g., graphene --
involves charged particles and Landau levels. The mechanism of magnetic catalysis is thus rather subtle and different from the mechanism
leading to magnetic catalysis in $d$=2+1 antiferromagnets where no charged particles are involved in its low-energy description. The fact
that the staggered magnetization grows in presence of a weak perpendicular magnetic field, is simply due to the suppression of quantum
fluctuations of the order parameter vector by the magnetic field. Still, according to the definition given in Ref.~\citep{ANT16}, we are
dealing with magnetic catalysis.

\subsection{Magnetization}

The low-temperature expansion of the magnetization,
\begin{equation}
\Sigma(T,H_s,H) = - \frac{\partial z(T,H_s,H)}{\partial H} \, ,
\end{equation}
takes the form\footnote{The coefficient ${\hat \sigma}_2$ can trivially be obtained from $z_{4b}$ given in Sec.~\ref{Evaluation}.}
\begin{eqnarray}
\label{magnetizationAF}
& & \Sigma(T,H_s,H) = \Sigma(0,H_s,H) + {\hat \sigma}_1 T + {\hat \sigma}_2 T^2 + {\cal O}(T^3) \, , \nonumber \\
& & \ {\hat \sigma}_1(T,H_s,H) = - H h^{I}_1 \, .
\end{eqnarray}
The free Bose gas contribution is proportional to one power of temperature, while the spin-wave interaction is contained in the
$T^2$-term.

\begin{figure}
\begin{center}
~~~~\hbox{
\includegraphics[width=7.5cm]{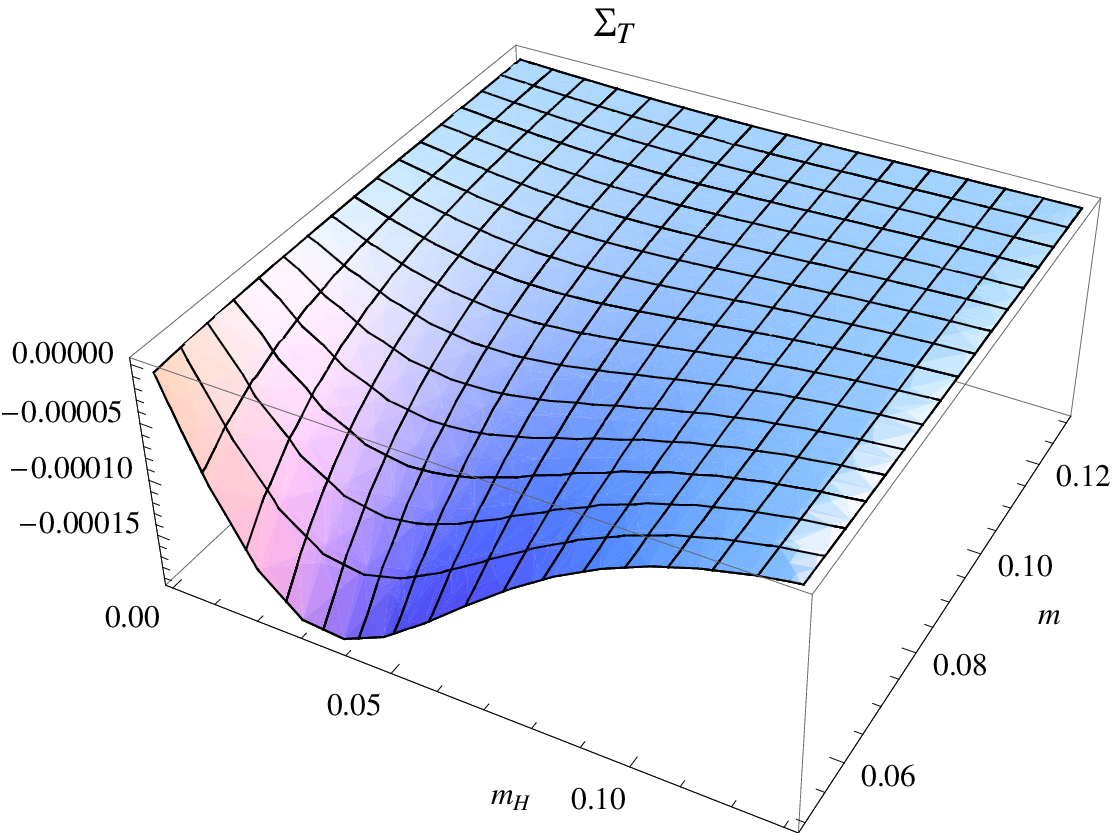}
\includegraphics[width=7.5cm]{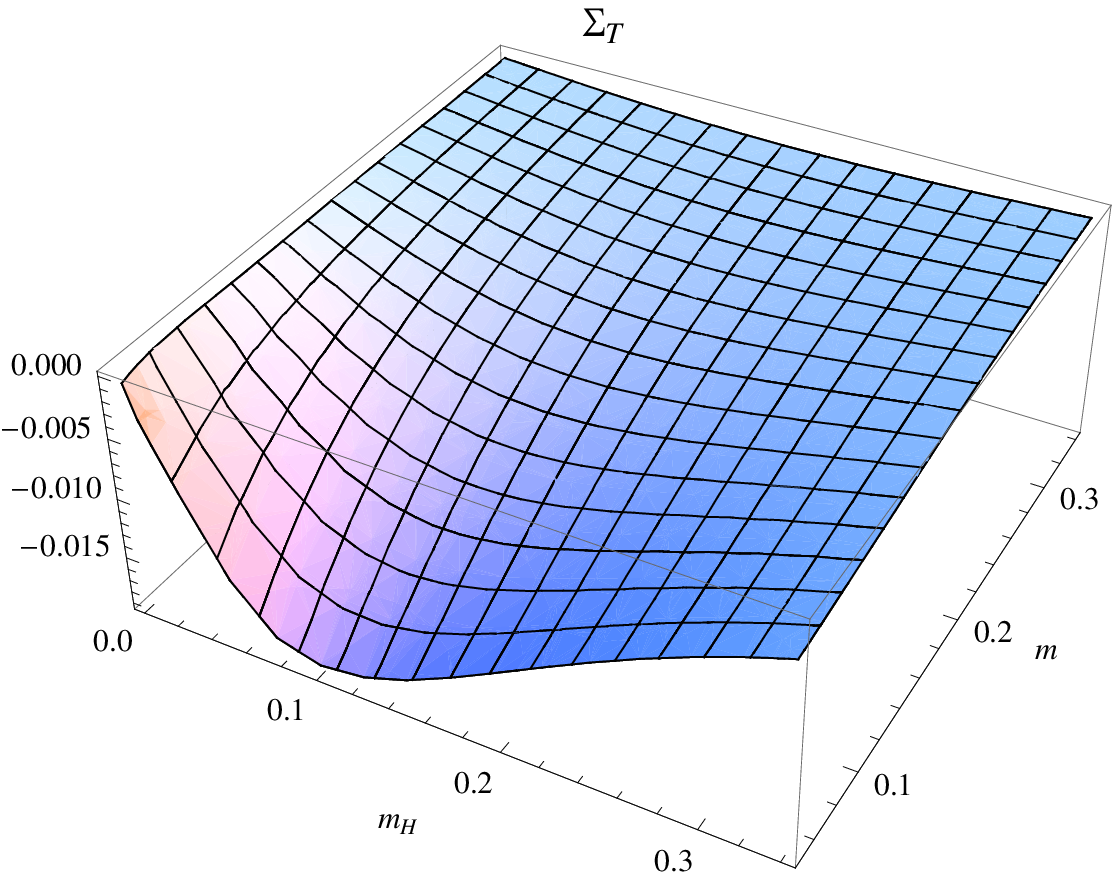}}
\end{center}
\caption{[Color online] Temperature-dependent part of the magnetization -- measured by $\Sigma_T(T,H_s,H)$ -- of $d$=2+1
antiferromagnets as a function of magnetic and staggered field at $T/2 \pi F^2 = 0.02$ (left) and $T/2 \pi F^2 = 0.1$ (right).}
\label{figure7}
\end{figure}

In Fig.~\ref{figure7}, for the temperatures $T/2 \pi F^2 = \{ 0.02, 0.1 \}$, we plot the total temperature-dependent part of the
magnetization 
\begin{equation}
\label{SigT}
\Sigma_T(T,H_s,H) = \frac{{\hat \sigma}_1 T + {\hat \sigma}_2 T^2}{F^4} \, .
\end{equation}
The quantity $\Sigma_T$ is negative in the entire parameter domain we consider. Negative $\Sigma_T$ means that the magnetization decreases
when we go from from $T$=0 to finite $T$ while keeping $H_s$ and $H$ fixed. This is what one would expect.

\begin{figure}
\begin{center}
~~~~\hbox{
\includegraphics[width=7.5cm]{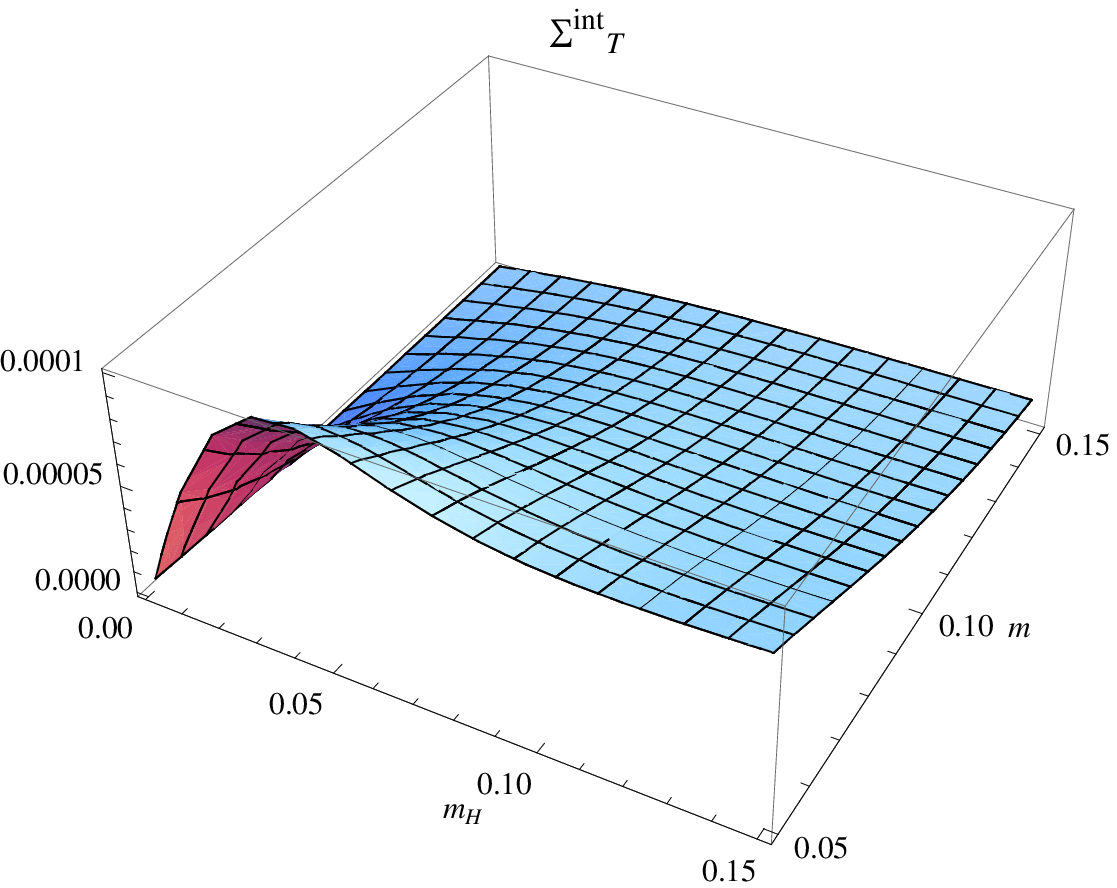}
\includegraphics[width=7.5cm]{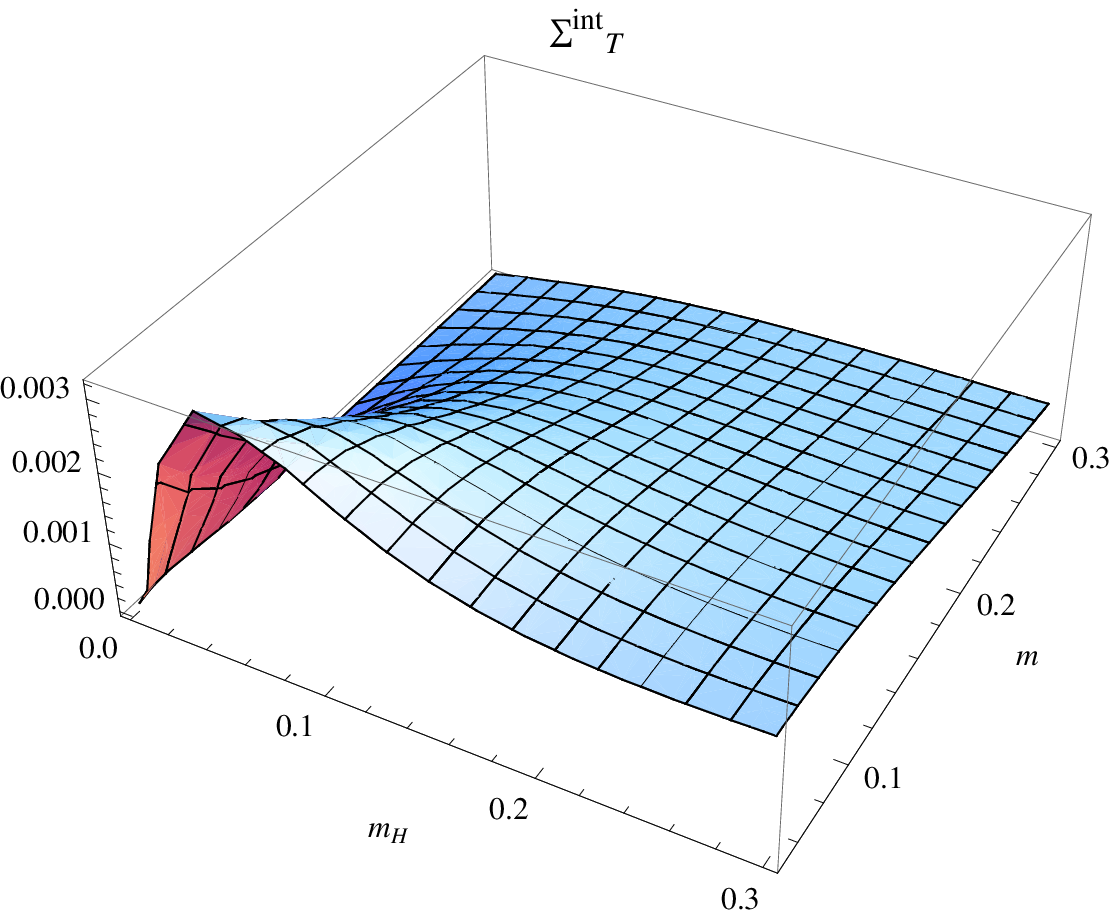}}
\end{center}
\caption{[Color online] Spin-wave interaction manifesting itself in the magnetization -- measured by $\Sigma^{int}_T(T,H_s,H)$ -- of
$d$=2+1 antiferromagnets as a function of magnetic and staggered field at the temperatures $T/2 \pi F^2 = 0.03$ (left) and
$T/2 \pi F^2 = 0.08$ (right).}
\label{figure8}
\end{figure}

Remarkably, the quantity 
\begin{equation}
\Sigma^{int}_T(T,H_s,H) = \frac{{\hat \sigma}_2 T^2}{F^4} \, ,
\end{equation}
that only takes into account the spin-wave interaction part, is positive as we illustrate in Fig.~\ref{figure8} that refers to the
temperatures $T/2 \pi F^2 = \{ 0.03, 0.08 \}$. Positive $\Sigma^{int}_T(T,H_s,H)$ means that if the temperature is raised from $T$=0 to
finite $T$ -- while keeping $H_s$ and $H$ fixed -- the magnetization grows due to the spin-wave interaction. This result appears to be
rather counterintuitive. But it is important to point out that we are dealing with weak effects originating from the spin-wave
interaction. The dominant behavior at finite temperature is given by the free Bose gas term. Indeed, the total temperature-dependent
magnetization (not just the interaction part), is strictly negative according to Fig.~\ref{figure7}.

Finally, at zero temperature, the magnetization amounts to
\begin{eqnarray}
\label{MagT0}
\frac{\Sigma(0,H_s,H)}{F^4} & = & 2 \pi \, m_H + \pi \, m_H \sqrt{m^2+m_H^2} + \pi m_H^3 + \frac{5 \pi}{8} \, m^2 \, m_H \nonumber \\
& &- \frac{\pi}{8} \, \frac{m^3 \, m_H}{\sqrt{m^2+m_H^2}} + 32 \pi^3 F^2 (e_1 + e_2) \, m_H^3 + 16 \pi^3 F^2 k_1 \, m^2 m_H \, , \nonumber \\
& & \hspace{-2.4cm} m = \frac{\sqrt{\Sigma_s H_s}}{2 \pi F^3} \, , \qquad m_H = \frac{H}{2 \pi F^2} \, .
\end{eqnarray}
Again, NLO effective constants -- $e_1, e_2, k_1$ -- show up in subleading corrections. If the magnetic field is switched off, the
magnetization tends to zero as it should,
\begin{equation}
\lim_{H \to 0}\Sigma(0,H_s,H) = 0 \, .
\end{equation}
In the limit $H_s \to 0$, the expansion in the magnetic field involves integer powers of $H$,
\begin{equation}
\label{MagnetizationinHzeroHs}
\Sigma(0,0,H) = F^2 H + \frac{H^2}{4 \pi} + \Big\{ 4 (e_1 + e_2) + \frac{1}{8 \pi^2 F^2} \Big\} \, H^3 + {\cal O}(H^4) \, .
\end{equation}
The leading contributions are positive, whereas the sign of $H^3$-term remains open. The leading terms, however, do not involve NLO
effective constants, such that the magnetization takes positive values in presence of the magnetic field. As one would expect, the
magnetization in the direction of the magnetic field no longer is zero, since the spins get tilted.

\section{Conclusions}
\label{conclusions}

We have considered the low-energy properties of antiferromagnetic films subjected to magnetic fields perpendicular to the staggered
magnetization order parameter. Within effective field theory we have systematically derived the low-temperature expansions for the free
energy density, pressure, order parameter, and magnetization.

In presence of a weak magnetic field, the spin-wave interaction in the pressure is repulsive, irrespective of the strength of the magnetic
and staggered field. The order parameter decreases due to the spin-wave interaction, when the temperature is raised from $T$=0 to finite
$T$ -- while keeping $H_s$ and $H$ fixed. Finally, the magnetization -- both at zero and finite temperature -- takes positive values: the
spins get tilted into the direction of the external perpendicular magnetic field.

At zero temperature, both the magnetization and staggered magnetization grow when a perpendicular magnetic field is applied. While this
behavior is expected for the magnetization, the enhancement of the order parameter in presence of the magnetic field comes rather
unexpectedly. It implies that the phenomenon of magnetic catalysis -- well-known in quantum chromodynamics, graphene and other condensed
matter systems -- also emerges in antiferromagnetic films.

\section*{Acknowledgments}
The author thanks J.\ O.\ Andersen, A.\ Auerbach, T.\ Brauner, H.\ Leutwyler, I.\ A.\ Shovkovy and R.\ R.\ P.\ Singh for correspondence.

\begin{appendix}

\section{Vertices with an Odd Number of Magnon Lines}
\label{appendixA}

Magnetic fields perpendicular to the staggered magnetization order parameter give rise to vertices that involve an {\it odd} number of
magnon lines. Explicitly, vertices with one magnon line originate from
\begin{equation}
\label{oneLine}
i F^2 H \partial_0 U^2 + 2 i k_1 \frac{\Sigma_s H_s}{F^2} \, H \partial_0 U^2 \, ,
\end{equation}
while vertices with three magnon lines are generated by
\begin{eqnarray}
\label{threeLines}
& & i F^2 H \Big\{ U^2 U^1 \partial_0 U^1 - \mbox{$ \frac{1}{2}$} \partial_0 U^2 U^1 U^1 + \mbox{$ \frac{1}{2}$} \partial_0 U^2 U^2 U^2
\Big\} \nonumber \\
& & + 2 i k_1 \frac{\Sigma_s H_s}{F^2} \, H \Big\{ U^1 \partial_0 U^1 U^2 - \partial_0 U^2 U^1 U^1 \Big\}
- 4 i(e_1 + e_2) H \partial_0 U^2 \partial_0 U^a \partial_0 U^a \nonumber \\
& & + i (e_1 + e_2) H^3 \Big\{ 4 U^1 \partial_0 U^1 U^2 + 2 U^2 \partial_0 U^2 U^2
- 6 \partial_0 U^2 U^1 U^1\Big\} \nonumber \\
& & - 4 i e_1 H \partial_0 U^2 \partial_r U^a \partial_r U^a - 4 i e_2 H \partial_r U^2 \partial_0 U^a \partial_r U^a \, .
\end{eqnarray}
Note that we only consider contributions from ${\cal L}^2_{eff}$ and ${\cal L}^4_{eff}$ -- higher-order pieces of the effective Lagrangian
also yield such vertices, but they do not contribute up to order $p^4$ in the partition function, as we argue below. The additional
Feynman diagrams that can be constructed from the expressions (\ref{oneLine}) and (\ref{threeLines}) are depicted in Fig.~\ref{figure2}.
According to (\ref{oneLine}), the line emitted (or absorbed) by a one-magnon vertex always corresponds to $U^2$. In case of a three-magnon
vertex, according to (\ref{threeLines}), we either have $U^2 U^2 U^2$ or $U^1 U^1 U^2$ -- in particular, three magnons of the same type
$U^1$ are never emitted or absorbed simultaneously.

An important observation that drastically reduces the number of additional Feynman graphs, is that the one-magnon vertices from
${\cal L}^2_{eff}$ and ${\cal L}^4_{eff}$ are irrelevant. In the evaluation of the partition function they lead to integrals of the form
\begin{equation}
\int d^3x \, d^3y \, d^3z \dots \, {(\partial_0)}^2 G^{I\!I}(x-y) \, {\cal F}(y,z, \dots) \, , \qquad x = (x_0,x_1,x_2) \, ,
\end{equation}
where $\partial_0$ is the Euclidean time derivative corresponding to the coordinate $x_0$. The function ${\cal F}(y,z, \dots)$, depending
on the topology of the diagram, may contain an arbitrary number of propagators that involve additional time and space derivatives. But the
point is that -- irrespective of the complexity of the diagram -- the integration over the coordinates of the first vertex, i.e.,
integration over the coordinates $x_0,x_1,x_2$ of the one-magnon vertex, is identically zero. One concludes that the relevant new diagrams
must involve vertices with at least three magnon lines.

This then leads to the two-loop diagrams $4c$ and $4d$ of Fig.~\ref{figure2}. Any other diagram that involves vertices with an odd number
of magnon lines is at least of order $p^5$, i.e., beyond the scope of the present study. Remarkably, the explicit evaluation of diagram
$4c$ yields zero,
\begin{equation}
z_{4c} =0 \, ,
\end{equation}
while the sunset diagram contributes with
\begin{equation}
\label{z4d}
z_{4d} = \frac{2 H^2}{F^2} \, \int_{\cal T} {\mbox{d}}^d x \, G^{I}(x) \, \partial_0 G^{I}(x) \, \partial_0 G^{I\!I}(x) \, .
\end{equation}
This integral over the torus ${\cal T} = {\cal R}^{d_s} \times S^1$, with circle $S^1$ defined as $- \beta / 2 \leq x_0 \leq \beta / 2$, is
divergent in the ultraviolet. In the subsequent appendix we show how to isolate the singularities and how to evaluate the finite pieces.

\section{Evaluation of the Sunset Diagram}
\label{appendixB}

In order to process the integral (\ref{z4d}), we decompose the thermal propagators $G^{I,I\!I}(x)$ as
\begin{equation}
G^{I,I\!I}(x) = \Delta^{I,I\!I}(x) + {\overline G}^{I,I\!I}(x) \, ,
\end{equation}
where the $\Delta^{I,I\!I}(x)$ are the zero-temperature propagators defined in Eq.~(\ref{regprop}). The integral then takes the form
\begin{eqnarray}
\label{sunsetDecomp}
& & \int_{\cal T} {\mbox{d}}^d x \, \Big(
{\overline G}^{I} \, \partial_0 {\overline G}^{I} \, \partial_0 {\overline G}^{I\!I}
+ \Delta^{I} \, \partial_0 {\overline G}^{I} \, \partial_0 {\overline G}^{I\!I}
+ {\overline G}^{I} \, \partial_0 \Delta^{I} \, \partial_0 {\overline G}^{I\!I}
+ {\overline G}^{I} \, \partial_0 {\overline G}^{I} \, \partial_0 \Delta^{I\!I} \nonumber \\
& & \hspace{0.6cm} + \Delta^{I} \, \partial_0 {\overline G}^{I} \, \partial_0 \Delta^{I\!I}
+ \Delta^{I} \, \partial_0 \Delta^{I} \, \partial_0 {\overline G}^{I\!I}
+ {\overline G}^{I} \, \partial_0 \Delta^{I} \, \partial_0 \Delta^{I\!I}
+ \Delta^{I} \, \partial_0 \Delta^{I} \, \partial_0 \Delta^{I\!I}
\Big) \, .
\end{eqnarray}
The first four integrals over the torus are convergent in $d \to 3$. The four remaining integrals that involve two or three
zero-temperature propagators, however, are singular in the limit $d \to 3$, and need to be considered in detail. Following
Ref.~\citep{GL89}, we cut out a sphere of radius $|S| \leq \beta/2$ around the origin and write the respective integrals over the
torus as 
\begin{eqnarray}
& & \int_{\cal T} {\mbox{d}}^d x \, \Delta^{I} \, \partial_0 {\overline G}^{I} \, \partial_0 \Delta^{I\!I}
= \int_{\cal S} {\mbox{d}}^d x \, \Delta^{I} \, \partial_0 {\overline G}^{I} \, \partial_0 \Delta^{I\!I}
+ \int_{{\cal T} \setminus {\cal S}} {\mbox{d}}^d x \, \Delta^{I} \, \partial_0 {\overline G}^{I} \, \partial_0 \Delta^{I\!I} \, , \nonumber \\
& & \int_{\cal T} {\mbox{d}}^d x \, \Delta^{I} \, \partial_0 \Delta^{I} \, \partial_0 {\overline G}^{I\!I}
= \int_{\cal S} {\mbox{d}}^d x \, \Delta^{I} \, \partial_0 \Delta^{I} \, \partial_0 {\overline G}^{I\!I}
+ \int_{{\cal T} \setminus {\cal S}} {\mbox{d}}^d x \, \Delta^{I} \, \partial_0 \Delta^{I} \, \partial_0 {\overline G}^{I\!I} \, , \nonumber \\
& & \int_{\cal T} {\mbox{d}}^d x \, {\overline G}^{I} \, \partial_0 \Delta^{I} \, \partial_0 \Delta^{I\!I}
= \int_{\cal S} {\mbox{d}}^d x \, {\overline G}^{I} \, \partial_0 \Delta^{I} \, \partial_0 \Delta^{I\!I}
+ \int_{{\cal T} \setminus {\cal S}} {\mbox{d}}^d x \, {\overline G}^{I} \, \partial_0 \Delta^{I} \, \partial_0 \Delta^{I\!I} \, , \nonumber \\
& & \int_{\cal T} {\mbox{d}}^d x \, \Delta^{I} \, \partial_0 \Delta^{I} \, \partial_0 \Delta^{I\!I}
= \int_{\cal S} {\mbox{d}}^d x \, \Delta^{I} \, \partial_0 \Delta^{I} \, \partial_0 \Delta^{I\!I}
+ \int_{{\cal T} \setminus {\cal S}} {\mbox{d}}^d x \, \Delta^{I} \, \partial_0 \Delta^{I} \, \partial_0 \Delta^{I\!I} \, .
\end{eqnarray}
The evaluation of the integrals over the complement of the torus ${\cal T} \setminus {\cal S}$ poses no problems in $d$=3. In the integral
over the sphere in line three, we subtract the piece $g^{I}_1={\overline G}^{I}|_{x=0}$,
\begin{equation}
{\overline G}^{I} \ \to \ {\overline G}^{I} - g^{I}_1 \, ,
\end{equation}
while in the integrals over the sphere in lines one and two, we perform the subtractions
\begin{equation}
\partial_0 {\overline G}^{I,I\!I} \ \to \ \partial_0 {\overline G}^{I,I\!I} - \, \partial^2_0 \, {\overline G}^{I,I\!I}|_{x=0} \, \times x_0
\, .
\end{equation}
Making use of 
\begin{equation}
\partial^2_0 \, {\overline G}^{I,I\!I}\!(x)|_{x=0} = g_0^{I,I\!I} + M^2_{I,I\!I} \, g_1^{I,I\!I} \qquad (d = 3) \, ,
\end{equation}
we end up with
\begin{eqnarray}
& & \int_{\cal S} {\mbox{d}}^d x \, \Delta^{I} \, \partial_0 {\overline G}^{I} \, \partial_0 \Delta^{I\!I}
= \int_{\cal S} {\mbox{d}}^d x \, \Delta^{I} \Big( \partial_0 {\overline G}^{I} - x_0 (g_0^{I} + M^2_{I} \, g_1^{I}) \Big) \, \partial_0
\Delta^{I\!I} \nonumber \\
& & \hspace{4.4cm} + \int_{\cal S} {\mbox{d}}^d x \, \Delta^{I} \, x_0 (g_0^{I} + M^2_{I} \, g_1^{I}) \, \partial_0 \Delta^{I\!I} \, ,
\nonumber \\
& & \int_{\cal S} {\mbox{d}}^d x \, \Delta^{I} \, \partial_0 \Delta^{I} \, \partial_0 {\overline G}^{I\!I}
= \int_{\cal S} {\mbox{d}}^d x \, \Delta^{I} \,\partial_0 \Delta^{I} \Big( \partial_0 {\overline G}^{I\!I} - x_0 (g_0^{I\!I} + M^2_{I\!I} \,
g_1^{I\!I}) \Big) \nonumber \\
& & \hspace{4.4cm} + \int_{\cal S} {\mbox{d}}^d x \, \Delta^{I} \partial_0 \Delta^{I} \, x_0 (g_0^{I\!I} + M^2_{I\!I} \, g_1^{I\!I}) \, ,
\nonumber \\
& & \int_{\cal S} {\mbox{d}}^d x \, {\overline G}^{I} \, \partial_0 \Delta^{I} \, \partial_0 \Delta^{I\!I}
= \int_{\cal S} {\mbox{d}}^d x \, \Big( {\overline G}^{I} - g^{I}_1 \Big) \partial_0 \Delta^{I} \, \partial_0 \Delta^{I\!I} \nonumber \\
& & \hspace{4.4cm} + \int_{\cal S} {\mbox{d}}^d x \, g^{I}_1 \, \partial_0 \Delta^{I} \, \partial_0 \Delta^{I\!I} \, .
\end{eqnarray}
The subtracted integrals over the sphere on the RHS are convergent in $d \to 3$. The second integrals on the RHS we decompose further as
\begin{eqnarray}
& & \int_{\cal S} {\mbox{d}}^d x \, \Delta^{I} \, x_0 (g_0^{I} + M^2_{I} \, g_1^{I}) \, \partial_0 \Delta^{I\!I}
= \int_{\cal R} {\mbox{d}}^d x \, \Delta^{I} \, x_0 (g_0^{I} + M^2_{I} \, g_1^{I}) \, \partial_0 \Delta^{I\!I} \nonumber \\
& & \hspace{4.9cm} - \int_{{\cal R} \setminus {\cal S}} {\mbox{d}}^d x \, \Delta^{I} \, x_0 (g_0^{I} + M^2_{I} \, g_1^{I}) \, \partial_0
\Delta^{I\!I} \, , \nonumber \\
& & \int_{\cal S} {\mbox{d}}^d x \, \Delta^{I} \partial_0 \Delta^{I} \, x_0 (g_0^{I\!I} + M^2_{I\!I} \, g_1^{I\!I})
= \int_{\cal R} {\mbox{d}}^d x \, \Delta^{I} \partial_0 \Delta^{I} \, x_0 (g_0^{I\!I} + M^2_{I\!I} \, g_1^{I\!I}) \nonumber \\
& & \hspace{4.9cm} - \int_{{\cal R} \setminus {\cal S}} {\mbox{d}}^d x \, \Delta^{I} \partial_0 \Delta^{I} \, x_0 (g_0^{I\!I} + M^2_{I\!I} \,
g_1^{I\!I}) \, , \nonumber \\
& & \int_{\cal S} {\mbox{d}}^d x \, g^{I}_1 \, \partial_0 \Delta^{I} \, \partial_0 \Delta^{I\!I}
= \int_{{\cal R}} {\mbox{d}}^d x \, g^{I}_1 \,\partial_0 \Delta^{I} \, \partial_0 \Delta^{I\!I}
- \int_{{\cal R} \setminus {\cal S}} {\mbox{d}}^d x \, g^{I}_1 \, \partial_0 \Delta^{I} \, \partial_0 \Delta^{I\!I} \, .
\end{eqnarray}
The integrals over the complement ${\cal R} \setminus {\cal S}$ are well-defined. The integrals over all Euclidean space are finite in
dimensional regularization in the limit $d \to 3$,
\begin{eqnarray}
& & \lim_{d \to 3} \, \int_{\cal R} {\mbox{d}}^d x \, \Delta^{I} \, x_0 (g_0^{I} + M^2_{I} \, g_1^{I}) \, \partial_0 \Delta^{I\!I}
= - \frac{M_{I}+ 2 M_{I\!I}}{12 \pi{(M_{I} + M_{I\!I})}^2} \, \Big( g^{I}_0 + M_{I}^2 g^{I}_1 \Big) \, , \nonumber \\
& & \lim_{d \to 3} \,\int_{\cal R} {\mbox{d}}^d x \, \Delta^{I} \partial_0 \Delta^{I} \, x_0 (g_0^{I\!I} + M^2_{I\!I} \, g_1^{I\!I})
= - \frac{1}{16 \pi M_{I}} \, \Big( g^{I\!I}_0 + M_{I\!I}^2 g^{I\!I}_1 \Big) \, , \nonumber \\
& & \lim_{d \to 3} \, \int_{{\cal R}} {\mbox{d}}^d x \, g^{I}_1 \, \partial_0 \Delta^{I} \, \partial_0 \Delta^{I\!I}
= - \frac{M_{I}^2 + M_{I} M_{I\!I} + M_{I\!I}^2}{12 \pi(M_{I} + M_{I\!I})} \, g^{I}_1 \, .
\end{eqnarray}

Finally, the last integral in Eq.~(\ref{sunsetDecomp}) that contains three zero-temperature propagators, is decomposed as
\begin{eqnarray}
& & \int_{\cal T} {\mbox{d}}^d x \, \Delta^{I} \, \partial_0 \Delta^{I} \, \partial_0 \Delta^{I\!I} 
= \int_{{\cal T} \setminus {\cal S}} {\mbox{d}}^d x \, \Delta^{I} \, \partial_0 \Delta^{I} \, \partial_0 \Delta^{I\!I}
+ \int_{\cal R} {\mbox{d}}^d x \, \Delta^{I} \, \partial_0 \Delta^{I} \, \partial_0 \Delta^{I\!I} \nonumber \\
& & \hspace{2.6cm} - \int_{{\cal R} \setminus {\cal S}} {\mbox{d}}^d x \, \Delta^{I} \, \partial_0 \Delta^{I} \, \partial_0 \Delta^{I\!I} \, .
\end{eqnarray}
The integrals over ${\cal T} \setminus {\cal S}$ and ${\cal R} \setminus {\cal S}$ are finite, but the integral over all Euclidean space
is singular in $d \to 3$. The corresponding counterterm $\cal C$,
\begin{equation}
{\cal C} = \int_{\cal R} {\mbox{d}}^d x \, \Delta^{I} \, \partial_0 \Delta^{I} \, \partial_0 \Delta^{I\!I} \, ,
\end{equation}
can be absorbed by NLO effective constants in $z_{4a}$, Eq.~(\ref{z2z4a}).

In conclusion, the first four integrals in the sunset contribution, Eq.~(\ref{sunsetDecomp}), are well-defined and can be evaluated
numerically in a straightforward manner, using the fact that the integrals are two-dimensional,
\begin{equation}
{\mbox{d}}^3 x = 2 \pi r dr dt \, .
\end{equation}
The evaluation of the remaining four integrals in Eq.~(\ref{sunsetDecomp}) is more subtle, but can be handled within dimensional
regularization using the method established in Ref.~\citep{GL89}. In the limit $d \to 3$, the final -- and finite -- representation for
the free energy density originating from the sunset diagram $4d$ reads
\begin{eqnarray}
\label{Jbar}
{\overline z_{4d}} & = & \frac{2 H^2}{F^2} \, \Bigg( {\int}_{\!\!\! {\cal T}} \! \! {\mbox{d}}^3 x \, T
+ {\int}_{\!\!\! {\cal T} \setminus {\cal S}} \! \! {\mbox{d}}^3 x \, U
+ {\int}_{\!\!\! {\cal S}} \! \! {\mbox{d}}^3 x \, V
- {\int}_{\!\!\! {\cal R} \setminus {\cal S}} \! \! {\mbox{d}}^3 x \, W + R \Bigg) \, , \nonumber \\
T & = & {\overline G}^{I} \, \partial_0 {\overline G}^{I} \, \partial_0 {\overline G}^{I\!I}
+ \Delta^{I} \, \partial_0 {\overline G}^{I} \, \partial_0 {\overline G}^{I\!I}
+ {\overline G}^{I} \, \partial_0 \Delta^{I} \, \partial_0 {\overline G}^{I\!I}
+ {\overline G}^{I} \, \partial_0 {\overline G}^{I} \, \partial_0 \Delta^{I\!I} \, , \nonumber \\
U & = & \Delta^{I} \, \partial_0 {\overline G}^{I} \, \partial_0 \Delta^{I\!I}
+ \Delta^{I} \, \partial_0 \Delta^{I} \, \partial_0 {\overline G}^{I\!I}
+ {\overline G}^{I} \, \partial_0 \Delta^{I} \, \partial_0 \Delta^{I\!I}
+ \Delta^{I} \, \partial_0 \Delta^{I} \, \partial_0 \Delta^{I\!I} \, , \nonumber \\
V & = & \Delta^{I} \Big( \partial_0 {\overline G}^{I} - x_0 (g_0^{I} + M^2_{I} \, g_1^{I}) \Big) \partial_0 \Delta^{I\!I}
+ \Delta^{I} \,\partial_0 \Delta^{I} \Big( \partial_0 {\overline G}^{I\!I} - x_0 (g_0^{I\!I} + M^2_{I\!I} \, g_1^{I\!I}) \Big) \nonumber \\
& & + \, \Big( {\overline G}^{I} - g^{I}_1 \Big) \partial_0 \Delta^{I} \, \partial_0 \Delta^{I\!I} \, , \nonumber \\
W & = & \Delta^{I} \, x_0 (g_0^{I} + M^2_{I} \, g_1^{I}) \, \partial_0 \Delta^{I\!I}
+ \Delta^{I} \partial_0 \Delta^{I} \, x_0 (g_0^{I\!I} + M^2_{I\!I} \, g_1^{I\!I})
+ g^{I}_1 \, \partial_0 \Delta^{I} \, \partial_0 \Delta^{I\!I} \nonumber \\
& & + \, \Delta^{I} \, \partial_0 \Delta^{I} \, \partial_0 \Delta^{I\!I} \, , \nonumber \\
R & = & - \frac{M_{I}+ 2 M_{I\!I}}{12 \pi{(M_{I} + M_{I\!I})}^2} \, \Big( g^{I}_0 + M_{I}^2 g^{I}_1 \Big)
- \frac{1}{16 \pi M_{I}} \, \Big( g^{I\!I}_0 + M_{I\!I}^2 g^{I\!I}_1 \Big)  \nonumber \\
& & - \frac{M_{I}^2 + M_{I} M_{I\!I} + M_{I\!I}^2}{12 \pi(M_{I} + M_{I\!I})} \, g^{I}_1 \, .
\end{eqnarray}
In a last step we collect all the above contributions in the dimensionless function $s(\sigma,\sigma_H)$,
\begin{equation}
\label{sunny}
{\overline z_{4d}} = \frac{2}{F^2} \, s(\sigma,\sigma_H) \, T^4 \, ,
\end{equation}
where the dimensionless ratios $\sigma$ and $\sigma_H$ are defined as
\begin{equation}
\sigma = \frac{\sqrt{\Sigma_s H_s}}{2 \pi F T} = m \frac{F^2}{T} \, , \qquad \sigma_H = \frac{H}{2 \pi T} = m_H \frac{F^2}{T} \, .
\end{equation}
A plot of $s(\sigma,\sigma_H)$ is provided in Fig.~\ref{figureSunny}. Note the final result for the function $s(\sigma,\sigma_H)$ must be
independent of the size of the sphere which is an academic invention. We have verified that different sizes of the sphere indeed lead to
the same $s(\sigma,\sigma_H)$.

\end{appendix}

\end{document}